\RequirePackage{fix-cm}
\documentclass[smallextended]{svjour3}       
\smartqed  
\usepackage{graphicx}
\usepackage{color}
\begin{document}

\title{Quantum gravity and the square of Bell operators}


\author{S. Aghababaei  \and H. Moradpour \and
        H. Shabani}


\institute{S. Aghababaei \at Department of Physics, Faculty of
Sciences, University of Sistan and Baluchestan, Zahedan, Iran.\\
\email{sarah.aghababaei@gmail.com}  \and H. Moradpour
(corresponding author) \at
             Research Institute for Astronomy and Astrophysics of Maragheh (RIAAM), University of Maragheh, P.O. Box 55136-553, Maragheh, Iran.\\
              \email{hn.moradpour@maragheh.ac.ir}  \and
            H. Shabani \at
              Department of Physics, Faculty of Sciences, University of Sistan and Baluchestan, Zahedan, Iran.\\ \email{h.shabani@phys.usb.ac.ir}}

\date{Received: date / Accepted: date}

\maketitle

\begin{abstract}
The Bell's inequality is a strong criterion to distinguish
classical and quantum mechanical aspects of reality. Its violation
is the net effect of the existence of non-locality in systems, an
advantage for quantum mechanics (QM) over classical physics. The
quantum mechanical world is under the control of the Heisenberg
uncertainty principle (HUP) that is generalized by quantum gravity
(QG) scenarios, called generalized uncertainty principle (GUP).
Here, the effects of GUP on the square of Bell operators of qubits
and qutrits are studied. The achievements claim that the violation
quality of the square of Bell inequalities may be a tool to get a
better understanding of the quantum features of gravity. In this
regard, it is obtained that the current accuracy of the
Stern-Gerlach experiments implies upper bounds on the values of
the GUP parameters.

\keywords{Quantum gravity \and Quantum non-locality}
\end{abstract}
\maketitle
\bigskip

\section{Introduction}

The pioneering work of Einstein-Podolski-Rosen (EPR) reveals the
non-local feature of physical realities \cite{EPR}, a property
which leads to the violation of Bell's inequality \cite{Bell}. One
cornerstone of the EPR thought experiment is the uncertainty
principle arisen from the non-commutativity of position ($x$) and
momentum ($p$) operators \cite{EPR,fran,oppen,alsina}. In order to
make this point more clear, let us consider the CHSH
(Clauser-Horne-Shimony) form of Bell's inequality \cite{CHSH}

\begin{eqnarray}\label{BCHSH}
\langle\hat B_{CHSH}\rangle&=&\langle\hat a\hat b+\hat a \hat b'+\hat a' \hat b- \hat a' \hat b'\rangle\nonumber\\
&=&\langle\hat a (\hat b+ \hat b')+ \hat a'(\hat b- \hat b')\rangle\nonumber\\
&=&\pm 2,
\end{eqnarray}

\noindent in which $\hat a, \hat a'$ and $\hat b, \hat b'$ are
operators, employed by \textbf{Alice} and \textbf{Bob},
respectively, with eigenvalues $\pm1$ that satisfy $\hat
a^{2}=\hat a'^{2}=\hat b^{2}=\hat b'^{2}=\hat 1$. It is also
obvious that operators of different particles commute with each
other. The square of Eq.~(\ref{BCHSH}) ($\equiv \hat
B_{CHSH}^{2}=\hat B_{CHSH}\cdot \hat B_{CHSH}$) is also obtained
as \cite{alsina}

\begin{eqnarray}\label{B^2QM}
\hat B_{CHSH}^{2}=4\hat1_a\otimes \hat1_b-[\hat a,\hat
a']\otimes[\hat b,\hat b'],
\end{eqnarray}

\noindent which exposes the vital role of commutators of employed
operators in $\hat B^{2}_{CHSH}$. Indeed, if operators commute
with each other, then we have $\hat B_{CHSH}^{2}=4\hat1_a\otimes
\hat1_b$ which happens whenever there is not any non-locality
\cite{alsina}. Otherwise for non-commutative variables, it is not
correct. For example, by considering the Pauli matrices, where
their commutator is $[\hat\sigma_j,
\hat\sigma_k]=2i\epsilon_{jkl}\hat\sigma_l$, the maximum violation
of Bell's inequality is $\langle\hat
B_{CHSH}\rangle=\sqrt{\langle\hat
B_{CHSH}^{2}\rangle}=\sqrt{8}=2\sqrt{2}$. This value violates the
inequality $\langle\hat B_{CHSH}\rangle\leq 2$, a strong signal to
non-locality. Here, $\epsilon_{jkl}$ denotes the antisymmetric
Levi-Civita tensor.


In fact, it is Heisenberg who has firstly noted that the
uncertainty relations constrain the knowledge stored in a quantum
mechanical system \cite{Heisenberg}. This property causes
non-locality and thus the violation of CHSH inequality
\cite{fran,cerec}. Finally, it is worth to mention that there are
numerous attempts to investigate the violation of CHSH inequality
theoretically \cite{CHSH,theCHSH} and experimentally
\cite{expCHSH}, and for a general review, one can see Refs.
\cite{Bellworks,Gizin}. It also seems that this inequality is
beneficial in the relativistic regimes
\cite{20,21,kim,fris,moradpour1,moradpour2}. In fact, the problem
of the effects of relativity on non-locality and entanglement goes
back to the pioneering work of Peres et al. \cite{pres}.

In the above case, including two-dimensional systems such as the
electron spin, the preassumption of $\pm1$ for the outcomes of
measurements is a vital condition. In the case of $d$-dimensional
systems such as those include orbital angular momentum, the story
becomes more complicated. In these cases, it seems that the
probabilistic versions of Bell's inequality are more suitable
\cite{pbell1,pbell2}. In fact, the measurement outcomes are not
essentially $\pm1$ which limits the applicability of the square of
the existent operational versions (in comparison with
Eq.~(\ref{B^2QM})) \cite{alsina}. For simplicity, let us consider
Bell's inequalities for three outcomes $\{0,1,2\}$, proposed by
Collins et al. \cite{pbell1},

\begin{eqnarray}
\hat C_{223}&=&2-3(\hat a^2+\hat b'^2)+\frac{3}{4}(\hat a\hat b+\hat a^2\hat b-\hat a'\hat b-\hat a'^2\hat b-\hat a\hat b^2\nonumber\\
&+&\hat a'\hat b^2+\hat a\hat b'-\hat a^2\hat b'+\hat a'\hat b'+\hat a'^2\hat b'+\hat a\hat b'^2-\hat a'\hat b'^2)\nonumber\\
&+&\frac{9}{4}(\hat a^2\hat b^2-\hat a'^2\hat b^2+\hat a^2\hat
b'^2+\hat a'^2\hat b'^2), \label{cglmpreal}
\end{eqnarray}

\noindent where the notation $\hat C_{223}$ denotes $2$ parties,
$2$ settings, and $3$ outcomes. The value is given by $\langle
\hat C_{223}\rangle=2(5-\gamma^2)/3\approx2\cdot9149$ for the
optimal state $|\psi\rangle=
(|00\rangle+\gamma|11\rangle+|22\rangle)/\sqrt{(2+\gamma^2)}$,
where $\gamma =(\sqrt{11}-\sqrt{3})/2 \approx 0\cdot7923$
\cite{valuc223}. The $z$ component of angular momentum can be
considered as an operator for which the states $|0\rangle,
|1\rangle, |2\rangle$ are correspond to $m=0$, $m=1$, and $m=2$,
respectively (three outcomes $\{0,1,2\}$). The square of the
operator $\hat C_{223}$ can easily be found as \cite{alsina}

\begin{eqnarray}
\hat C^{2}_{223}=3+(1+\{\{\hat a,\hat a'\}\})(1+\{\{\hat b,\hat
b'\}\}), \label{C2}
\end{eqnarray}

\noindent in which $\{\{\hat a, \hat a'\}\}$ denotes the complex
anti-commutator $\{\{\hat a, \hat a'\}\}= \hat a\hat
a'^{\dagger}+\hat a'\hat a^{\dagger}$. In summary, all of the
above cases authenticate the role of the commutation relations, or
equally the uncertainty principles, in emerging the non-locality.

%
%
%
%

The quality of the violation of Bell inequality in the presence of
a gravitational field is firstly studied in Ref.~\cite{tera}.
Additionally, the effects of curved spacetimes and also the
presence of acceleration on entanglement and non-locality are
investigated in various articles such as
Refs.~\cite{fu,al,ma,le,fun,smith,alsing,shi,ball,ver,refn}. In
this regard, it is worthwhile to mention that the quantum features
of gravity propose modified forms of ordinary HUP
\cite{GUP,GUPworks,gupclass} and signal us to a minimal length
\cite{minimallength}. Such modified forms are also proposed in
optics \cite{opticGUP}.

Therefore, it is expected that QG affects our understanding of
non-locality which may even give us a way to test the quantum
gravity scenarios. There are several phenomenological studies on
GUP which leads to modifications in several areas of QM
\cite{GUPworks}. Due to GUP, it is proposed that the commutation
relations such as angular momentum operators \cite{angularGUP} and
spin algebra \cite{angularGUP,GUPspin} are modified which may give
us a possibility to verify the Planck scale effects in low energy
quantum systems \cite{GUPworks,gupclass}.

The aim of paper is to address the effects of quantum aspects of
gravity (GUP) on the square of Bell inequalities for the systems
including observables with two, and three outcomes. In the next
section, we provide an introductory note on GUP, and its
implications on the algebra of angular momentum, studied in
Ref.~\cite{angularGUP}. The square of Bell inequality for
spin-$1/2$ systems in the presence of GUP shall be investigated in
the third section. Three-level systems together with a conclusion
are also presented in the subsequent sections, respectively.

\section{GUP formalism}

Whenever the quantum features of gravity are considered, the
generalized coordinates $\hat\mathbf{X}$ and $\hat\mathbf{P}$
emerge instead of canonical coordinates $\hat\mathbf{x}$,
$\hat\mathbf{p}$, and in a quadratic model of GUP, proposed in
Ref.~\cite{GUP}, the HUP modification takes a momentum dependent
quadratic term as

\begin{eqnarray}
\Delta \hat X \Delta \hat P\geq \frac{\hbar}{2}\left(1+ \beta
\Delta \hat P^{2}\right), \label{GUPform}
\end{eqnarray}

\noindent where $\beta$ denotes the GUP parameter and a
fundamental minimal length is obtained as $\Delta X_{min} = \hbar
\sqrt{\beta}$, which is of the order of Planck's length ($l_{p} =
\sqrt{\hbar G / c^{3}}$). The above GUP is obtained by using the
modified Heisenberg algebra \cite{GUP,GUPworks,gupclass}

\begin{eqnarray}
[\hat X,\hat P] = i\hbar (1+\beta\hat P^{2}). \label{GUPform2}
\end{eqnarray}

The quantum mechanical commutators are replaced by the Poisson
bracket (PB) for corresponding classical variables by considering
the classical limit (i.e., $\hbar \rightarrow 0$). It means that
\cite{gupclass}

\begin{eqnarray}
\frac{1}{i\hbar}[\hat X,\hat P] \rightarrow \left \{ \hat X, \hat
P \right \}_{PB},
\end{eqnarray}

\noindent and thus \cite{gupclass}

\begin{eqnarray}
\left \{ \hat X, \hat P \right \}_{PB} = 1+\beta \hat P^{2}.
\label{poisonGUP}
\end{eqnarray}

To construct a general framework to study the GUP effects, we
introduce a representation, called coordinate representation, in
the form of

\begin{eqnarray}
\hat X&=&\hat x,\nonumber\\
\hat P&=&\hat p\left(1+\beta \hat p^2\right), \label{correp}
\end{eqnarray}

\noindent where $\hat x=(\hat x,\hat y,\hat z)$ and $\hat p=(\hat
p_x,\hat p_y,\hat p_z)$ represent the position and momenta
operators in QM, respectively. This representation can connect the
$(\hat\mathbf{X},\hat\mathbf{P})$ space near Planck scale (QG) to
the $(\hat\mathbf{x},\hat\mathbf{p})$ space of QM, employed in the
various works \cite{GUPworks,gupclass,rep}.

It is shown that Eq.~(\ref{GUPform2}) modifies the angular
momentum algebra, including the orbital angular momentum and spin
algebra \cite{angularGUP,GUPspin}.
In the presence of GUP, using Eq.~(\ref{correp}) one can easily
find out the modified algebra of orbital angular momentum (spin)
as \cite{angularGUP}

\begin{eqnarray}
&&[\hat L_i, \hat L_j]=i\epsilon_{ijk}\hat L_k(1+\beta\hat
P^{2}). \label{GUPr}
\end{eqnarray}

\noindent Here, $\hat L=\hat X\times \hat P$. By bearing
Eq.~(\ref{correp}) and $\hat\mathcal{L}=\hat x\times \hat p$ in
mind, one can use the above result to find out $\hat L=\hat
X\times \hat P=\hat \mathcal{L}(1+\beta\hat p^{2})$, and thus
$[\hat L_i, \hat L_j]=i\epsilon_{ijk}\hat\mathcal{L}_k(1+\beta\hat
p^2){(1+\beta\hat p^{2}(1+\beta\hat p^2)^2)}\approx
i\epsilon_{ijk}\hat\mathcal{L}_k (1+2\beta\hat p^2)$ where the
latter is written by considering only the terms of the order of
$\beta$, and for simplicity, we set $\hbar=1$. It is obvious that
the standard commutation relation of angular momentum is recovered
as $\beta\longrightarrow0$.
\section{Two-qubit systems}

Now, considering the set of operators $\{\hat A, \hat A', \hat B,
\hat B'\}$, with eigenvalues $\pm1$ satisfying the condition $\hat
A^{2}=\hat A'^{2}=\hat B^{2}=\hat B'^{2}=\hat 1$. But, here, they
obey relation~(\ref{GUPr}) instead of purely quantum mechanical
commutator (obtained for $\beta=0$), and by rewriting Bell's
operator in Eq.~(\ref{BCHSH}) with these operators, one finds

\begin{eqnarray}
\hat B_{GUP}=\hat A\otimes \hat B+\hat A\otimes \hat B'+\hat
A'\otimes \hat B-\hat A'\otimes \hat B'.
\end{eqnarray}

\noindent Therefore, the corresponding square of Bell's operator
is obtained as

\begin{eqnarray}
\hat B^{2}_{GUP}&=&\hat A^{2}\otimes \hat B^{2}+\hat A^{2}\otimes \hat B'^{2}+\hat A'^{2}\otimes \hat B^{2}+\hat A'^{2}\otimes \hat B'^{2}\nonumber\\
&+&(\hat A^{2}-\hat A'^{2})\otimes \{\hat B, \hat B'\}_{PB}\nonumber\\
&+&\{\hat A, \hat A'\}_{PB}\otimes (\hat B^{2}-\hat B'^{2})\nonumber\\
&-&[\hat A, \hat A']\otimes [\hat B, \hat B'],
\end{eqnarray}

\noindent that finally leads to

\begin{eqnarray}\label{fs} \hat
B^{2}_{GUP}&=&4\hat1_A\otimes \hat1_B-[\hat A, \hat A']\otimes
[\hat B, \hat B'],
\end{eqnarray}

\noindent similar to Eq.~(\ref{B^2QM}). In order to get more
detailed analysis, let us firstly consider a quantum state
$|\varphi\rangle$, which includes spin and momentum information of
the system, and the unit vectors
$\overrightarrow{a}=(a_x,a_y,a_z)$ and
$\overrightarrow{a'}=(a'_x,a'_y,a'_z)$ for the directions of
operators $\hat A$ and $\hat A'$, respectively, and
$\overrightarrow{b}=(b_x,b_y,b_z)$ and
$\overrightarrow{b'}=(b'_x,b'_y,b'_z)$ for $\hat B$ and $\hat B'$,
respectively. Based on Eq.~(\ref{GUPr}) and
Ref.~\cite{angularGUP}, we have $\hat S=\hat s(1+\beta\hat p^2)$,
where $\hat s$ is the spin operator in the pure quantum mechanical
regime, and $\hat S_i$ (spin operator in the presence of minimal
length with eigenvalues $\lambda=\pm\frac{1}{2}(1+\beta p^2)$)
obey algebra~(\ref{GUPr}). Therefore, defining the operator $\hat
O\equiv\frac{\hat S\cdot \overrightarrow{n}}{|\lambda|}$, where
$\overrightarrow{n}=\{\overrightarrow{a},\overrightarrow{a'},\overrightarrow{b},\overrightarrow{b'}\}$,
we can get operators $\hat A, \hat A', \hat B, \hat B'$,
respectively. In this manner, bearing in mind that spin and
momentum commute with each other, $\big[\hat p_i,\hat P_i\big]=0$,
and by using Eq.~(\ref{GUPr}), we easily find

\begin{eqnarray}\label{fs1} \langle\hat
B^{2}_{GUP}\rangle=4+\frac{(1+\beta P_{1}^{2})(1+\beta
P_{2}^{2})}{\lambda_1^2\lambda_2^2}
\langle(\sum_{i,j}a_ia'_j\epsilon_{ijk}\hat
S_k^1)(\sum_{l,m}b_lb'_m\epsilon_{lmn}\hat S_n^2)\rangle,
\end{eqnarray}

\noindent where $ P_i$ and $\hat S^i$ denote the momentum and the
spin operators of the $i$-th particle in the QG regime,
respectively. Moreover, $\lambda_i=\frac{1}{2}(1+\beta p_i^2)$
(corresponding to the $i$-th particle), one finally obtains

\begin{eqnarray}\label{fs2} \langle\hat
B^{2}_{GUP}\rangle&\simeq&4+\frac{(1+\beta P^2)^2}{\lambda^2}
\langle(\sum_{i,j}a_ia'_j\epsilon_{ijk}\hat
\sigma_k^1)(\sum_{l,m}b_lb'_m\epsilon_{lmn}\hat
\sigma_n^2)\rangle,
\end{eqnarray}

\noindent in which we considered $p_a=p_b\equiv p$ (particles have
the same momentum), and $\hat s_i=\frac{1}{2}\hat\sigma_i$ (Pauli
matrices) has also been used. Now, bearing Eq.~(\ref{correp}) in
mind, since $\frac{(1+\beta
P^2)^2}{\lambda^2}\approx4\frac{1+2\beta P^2}{1+2\beta
p^2}\approx4[1+4\beta^{2}p^4]$, we have $\langle\hat
B^{2}_{GUP}\rangle\simeq\langle\hat B_{CHSH}^{2}\rangle$ up to the
first order of $\beta$, and

\begin{eqnarray}\label{fs22}\langle\hat
B^{2}_{GUP}\rangle &\simeq& \langle\hat B_{CHSH}^{2}\rangle+
16\beta^2 p^4
\langle(\sum_{i,j}a_ia'_j\epsilon_{ijk}\hat
\sigma_k^1)(\sum_{l,m}b_lb'_m\epsilon_{lmn}\hat
\sigma_n^2)\rangle,
\end{eqnarray}

\noindent up to the second order of $\beta$. Hence, the existence
of non-zero minimum length affects the square of Bell inequality,
due to the fact that the commutation relations are modified in the
presence of a non-zero minimum length. In general, for quantum
states like the Bell states, the maximum value of $\langle\hat
B_{CHSH}^{2}\rangle$ is achieved, as we have
$\langle(\sum_{i,j}a_ia'_j\epsilon_{ijk}\hat
\sigma_k^1)(\sum_{l,m}b_lb'_m\epsilon_{lmn}\hat
\sigma_n^2)\rangle=1$, that finally leads to $\langle\hat
B^{2}_{GUP}\rangle\simeq8+16\beta^2 p^4$. In summary, when the
quantum features of gravity become non-ignorable, $\hat
B^{2}_{GUP}$ is the true square of the Bell operator. The
difference between $\hat B^{2}_{GUP}$ and $\hat B^2_{CHSH}$ is
mathematically due to the effects of QG on the commutation
relations. It means that we may get a better understanding of spin
in high energy physics by using much more accurate apparatus in
the future, a result in line with previous studies (see~\cite{njp}
and references therein).

Here, it is worthwhile to focus on a more general GUP framework
\cite{Ali:2011fa} in which

\begin{eqnarray}
\hat P&=&\hat p\left(1+f(\hat p)\right),
\end{eqnarray}

\noindent where $f(\hat p)=\alpha \hat p+\beta \hat p^{2}$,
$\alpha$ and $\beta$ also denote the corresponding GUP parameters
in the linear and quadratic terms, respectively. By considering
this general form and following the approach of
Ref.~\cite{angularGUP}, the alternative of angular momentum
algebra described in Eq.~(\ref{GUPr}) is obtained
as~\cite{angularGUP}

\begin{eqnarray}
    &&[\hat L_i, \hat L_j]=i\epsilon_{ijk}\hat L_k(1+f(\hat P)), \label{GUPr2}
\end{eqnarray}

\noindent that finally leads to

\begin{eqnarray}
    \label{general} \langle\hat
    B^{2}_{GUP}\rangle&\simeq&4+4\frac{(1+ f(P))^2}{(1+f(p))^2}
    \langle(\sum_{i,j}a_ia'_j\epsilon_{ijk}\hat
    \sigma_k^1)(\sum_{l,m}b_lb'_m\epsilon_{lmn}\hat \sigma_n^2)\rangle,\nonumber\\
    &\simeq&\langle\hat B_{CHSH}^{2}\rangle\\\nonumber
    &+&\bigg(8\alpha^{2} p^{2}+16\beta^{2} p^{4}+\cdots\bigg)\langle(\sum_{i,j}a_ia'_j\epsilon_{ijk}\hat
    \sigma_k^1)(\sum_{l,m}b_lb'_m\epsilon_{lmn}\hat
    \sigma_n^2)\rangle.
\end{eqnarray}

\noindent Therefore, up to the leading order, one easily finds
$\langle\hat B^{2}_{GUP}\rangle\simeq 8+8\alpha^{2}
p^{2}+16\beta^2 p^4$, for the Bell states leading to
$\langle(\sum_{i,j}a_ia'_j\epsilon_{ijk}\hat
    \sigma_k^1)(\sum_{l,m}b_lb'_m\epsilon_{lmn}\hat
    \sigma_n^2)\rangle=1$.

Now, to examine Eq.~(\ref{general}), let us consider the
Stern-Gerlach experiment described in \cite{SG,SG2021}. This
experiment includes atoms of $ p^{2}=2.8 \times 10^{-26}
(\frac{\rm kg.m}{s})^{2}$ combined with $|\langle\frac{\hat
B^{2}_{GUP}-\hat B^{2}_{CHSH}}{\hat B^{2}_{CHSH}}\rangle|\simeq
\alpha^{2} p^{2}+2\beta^{2} p^{4}$ to provide two upper bounds on
$\alpha$ ($\beta$) as $\alpha_{0}\ll 10^{13}$
($\beta_0\ll10^{26}$), and $\alpha_{0}\ll 10^{11}$
($\beta_0\ll10^{24}$) for the splitting accuracies $ 10^{-1}$
\cite{SG}, and $10^{-3}$ \cite{SG2021}, respectively. Here,
$\alpha_0\equiv\alpha M_{p}c^{2}$ ($\beta_0\equiv\beta
M_{p}^{2}c^{4}$), $M_p$ denotes the Planck mass, and the obtained
upper bounds are well comparable with previous reports
\cite{GUPworks,angularGUP,bounds}.

\section{Two qutrits}

The corresponding operator for three outcomes, and its square,
introduced in Eq.~(\ref{cglmpreal}), are rewritten in the GUP
framework as

\begin{eqnarray}
(\hat C_{223})_{GUP}&=&2-3(\hat A^2+\hat B'^2)\nonumber\\
&+&\frac{3}{4}(\hat A\hat B+\hat A^2\hat B-\hat A'\hat B-\hat
A'^2\hat B-\hat A\hat B^2+
\hat A'\hat B^2\nonumber\\&+&\hat A\hat B'-\hat A^2\hat B'+\hat A'\hat B'+\hat A'^2\hat B'+\hat A\hat B'^2-\hat A'\hat B'^2)\nonumber\\
&+&\frac{9}{4}(\hat A^2\hat B^2-\hat A'^2\hat B^2+\hat A^2\hat
B'^2+\hat A'^2\hat B'^2),
\end{eqnarray}

\noindent and

%
    \begin{eqnarray}
        (\hat C^{2}_{223})_{GUP}&=&3+(1+\{\{\hat A,\hat A'\}\})(1+\{\{\hat B,\hat
        B'\}\}),
        \label{c2GUP}
\end{eqnarray}

\noindent respectively. Here, the operator $\hat O\in\{\hat A,\hat
A',\hat B,\hat B'\}$ contains three outcomes $\{0,1,2\}$. For
example, consider a particle with quantum mechanical momentum $p$
whose angular momentum meets algebra~(\ref{GUPr}), the operator
$\frac{\hat L_z}{1+\beta p^2}$ leads to outcomes $0,1,2$ for
states $|0\rangle, |1\rangle, |2\rangle$, respectively. Now,
following Ref.~\cite{angularGUP}, one can find $\{\{\hat L_i,\hat
L_j'\}\}=(1+\beta \hat P^{2})^{2}\{\{\hat l_i,\hat l_j'\}\}$, and
hence, $\{\{\hat O_i,\hat O_j'\}\}=\frac{(1+\beta \hat
P^{2})^{2}}{(1+\beta p^2)^2}\{\{\hat l_i,\hat l_j'\}\}$. Bearing
all of these points in mind, after some calculations, we finally
get

%
%

\begin{eqnarray}\label{qu}
        \langle(\hat C^{2}_{223})_{GUP}\rangle
        &=& \langle\hat C^{2}_{223}\rangle+4\beta^{2}p^{4}\underbrace{\langle\{\{\hat a,\hat
        a'\}\}+\{\{\hat b,\hat
        b'\}\}+2\{\{\hat a,\hat
        a'\}\}\{\{\hat b,\hat
        b'\}\}\rangle}_{\langle\hat G\rangle}\nonumber\\
        &+&\mathcal{O}(\beta^{3}),
    \end{eqnarray}

\noindent whenever all particles have the same momentum. Here, it
is again clear that $\langle(\hat
C^{2}_{223})_{GUP}\rangle\rightarrow\langle\hat
C^{2}_{223}\rangle$ as $\beta\rightarrow0$. Although, the
usefulness of $\langle\hat C_{223}^2\rangle$ in studying the
non-locality is challenging \cite{valuc223}, Eq.~(\ref{qu})
clearly shows that $\langle(\hat C^{2}_{223})_{GUP}\rangle$
differs from $\langle\hat C_{223}^2\rangle$ meaning that this
operator may be employed to investigate the predictions of QG. As
an example, consider $\hat a=\hat a'=\hat l_z$, $\hat b=\hat
b'=\hat l_z$, and the state $|\psi\rangle=
\frac{|00\rangle+|11\rangle+|22\rangle}{\sqrt{3}}|pp\rangle$ where
two particles have the same momentum $p$ (and thus $P$). In this
manner, $\{\{\hat l_z,\hat l_z\}\}=2\hat l_z^2$ leading to
$\langle\hat C_{223}^2\rangle\simeq32\cdot7$, and $\langle\hat
G\rangle=52$, and we finally have $\langle\hat
C_{223}^2\rangle\neq\langle(\hat C^{2}_{223})_{GUP}\rangle$.

\section{Conclusion}

The belief that the world is non-local comes from the amazing EPR
paper \cite{EPR} that motivated Bell to introduce his inequality.
Indeed, a cornerstone of the EPR argument is the role of HUP
(commutation relations) in the emergence of non-locality which
emerges in the square of Bell operators. On the other hand, it is
believed that the quantum aspects of gravity affect HUP, and
according to this proposal, we tried to shed light on the relation
between non-locality and QG. Related experiments and studies may
help us achieve a better understanding of gravity, and its
relation with non-locality, and quantum mechanics. Indeed, the
hopes to test the QG scenarios via studying its relation with
non-locality can be strengthened by increasing the accuracy of
related experiments such as the Stern-Gerlach apparatus.

\section*{Acknowledgment}
The authors would like to appreciate the anonymous referees for
providing useful comments and suggestions that helped us improve
the manuscript.

\end{document}